\documentclass[aps,twocolumn,preprintnumbers,amsmath,amssymb,nofootinbib,superscriptaddress,notitlepage]{revtex4-1}

\usepackage{slashed}
\usepackage{soul}
\usepackage{mathrsfs,bm}
\usepackage{txfonts}
\usepackage{amssymb}
\usepackage{indentfirst}
\usepackage{graphicx,booktabs}
\usepackage{multirow}
\usepackage{overpic}
\usepackage{color}
\usepackage{amssymb}

\begin{document}
\title{Heavy{-}Quark Symmetry Partners of the {$\bm{P_c(4450)}$} Pentaquark}

\author{Ming-Zhu Liu}

\author{Fang-Zheng Peng}
\affiliation{School of Physics and
Nuclear Energy Engineering,  Beihang University, Beijing 100191, China}

\author{Mario {S\'anchez S\'anchez}}
\affiliation{Centre d'\'Etudes Nucl\'eaires, CNRS/IN2P3, Universit\'e de Bordeaux, 33175 Gradignan, France}

\author{Manuel {Pavon} Valderrama}\email{mpavon@buaa.edu.cn}
\affiliation{School of Physics and
Nuclear Energy Engineering,  Beihang University, Beijing 100191, China}
\affiliation{International Research Center for Nuclei and Particles
  in the Cosmos \&
  Beijing Key Laboratory of Advanced Nuclear Materials and Physics,
  Beihang University, Beijing 100191, China}

\date{\today}
\begin{abstract}
  The spectrum of heavy{-}hadron molecules is constrained by
  heavy{-}quark symmetry in its different manifestations.
  Heavy{-}quark spin symmetry for instance connects the properties of
  the ground and excited states of heavy hadrons, while
  heavy{-}antiquark-diquark symmetry connects
  the properties of heavy antimesons ($\bar{D}$,{ }$\bar{D}^*$)
  and doubly heavy baryons ($\Xi_{cc}$, $\Xi_{cc}^*$).
  A prediction of these symmetries is that if the $P_c(4450)$ is indeed a
  $\bar{D}^* \Sigma_c$ bound state, then there should be a series of
  $\bar{D}^* \Sigma_c^*$, $\Xi_{cc} \Sigma_c$, $\Xi_{cc}^* \Sigma_c$,
  $\Xi_{cc} \Sigma_c^*$ and $\Xi_{cc}^* \Sigma_c^*$ partners.
  The concrete application of heavy{-}quark spin symmetry
  indicates that,
  if the $P_c(4450)$ is a $\frac{3}{2}^{-}$ $\bar{D}^* \Sigma_c$ molecule,
  the existence of a $\frac{5}{2}^{-}$ $\bar{D}^* \Sigma_c^*$ partner 
  with{ }similar binding energy --- which
  we call $P_c(4515)$, given its expected mass --- is likely.
  Conversely, the application of heavy{-}antiquark-diquark symmetry indicates
  that the $0^+$ $\Xi_{cc} \Sigma_c$, $1^+$ $\Xi_{cc} \Sigma_c^*$,
  $2^+$ $\Xi_{cc}^* \Sigma_c$ and $3^+$ $\Xi_{cc}^* \Sigma_c^*$
  molecules are likely to bind too, with binding energies
  in the $20-30\,{\rm MeV}$ range.
\end{abstract}

\pacs{13.60.Le, 12.39.Mk,13.25.Jx}

\maketitle

Exotic hadrons --- hadrons that are neither a quark-antiquark or
a three-quark state --- are an interesting window
into low-energy QCD dynamics.
From a theoretical perspective the {simplest} type of exotic hadron{s}
are hadronic molecules, which are bound states of two or more hadrons.
They were theorized decades ago~\cite{Voloshin:1976ap,DeRujula:1976zlg}
on the analogy of how the nuclear forces among nucleons generate
the deuteron and other nuclei.
The discovery of the $X(3872)$~\cite{Choi:2003ue} provided the first
solid candidate for a hadronic molecule and suggested that
the early speculations about
their existence~\cite{Tornqvist:1993ng,Ericson:1993wy}
were on the right track.
The proximity of the $X(3872)$ to the open{-}charm threshold $D^{*0} \bar{D}^0$
provides circumstantial evidence that the $X(3872)$ is molecular~\cite{Tornqvist:2003na,Voloshin:2003nt,Braaten:2003he},
while the isospin{-}breaking decays into $J/\psi\,2\pi$ and
$J/\psi\,3\pi$~\cite{Choi:2011fc}
represent a stronger case for its molecular nature~\cite{Gamermann:2009fv,Gamermann:2009uq,Hanhart:2011tn}.
The most stringent test of the $X(3872)$ nature will eventually be provided
by its decays into $D^0 \bar{D}^{0} \gamma$ and
$D^0 \bar{D}^{0} \pi^0$~\cite{Voloshin:2003nt,Fleming:2007rp,Guo:2014hqa},
but as for now the detailed experimental information required
about them is not available.
Other molecular candidates include
the $Z_b(10610)$ and $Z_b(10650)$~\cite{Belle:2011aa,Garmash:2014dhx},
the $Z_c(3900)$~\cite{Ablikim:2013mio, Liu:2013dau}
and $Z_c(4020)$~\cite{Ablikim:2013wzq, Ablikim:2014dxl},
see Ref.~\cite{Guo:2017jvc} for a recent review.
Recently a narrow pentaquark-like resonance, the $P_c(4450)$,
was discovered by the LHCb~\cite{Aaij:2015tga}, which has been assumed to be a $\bar{D}^* \Sigma_c$~\cite{Karliner:2015ina,Roca:2015dva,Xiao:2015fia,Burns:2015dwa,Geng:2017hxc}, a $\bar{D}^* \Sigma_c^*$~\cite{Chen:2015loa,Chen:2015moa},
{or a $\chi_{c1}p$ molecule}~\cite{Meissner:2015mza}.
Besides the molecular hypothesis, there are are other competing
explanations for the $P_c(4450)$: a genuine pentaquark~\cite{Diakonov:1997mm,Jaffe:2003sg,Yuan:2012wz,Maiani:2015vwa,Lebed:2015tna,Li:2015gta},
a threshold effect~\cite{Guo:2015umn,Liu:2015fea} (see~\cite{Bayar:2016ftu} for a detailed discussion), baryocharmonium~\cite{Kubarovsky:2015aaa} or other more exotic possibilities~\cite{Mironov:2015ica,Scoccola:2015nia}.

Hadronic molecules have a high degree of symmetry.
If the hadrons conforming a molecule contain light quarks, 
chiral and SU(3)-flavour symmetries will strongly constrain
the interactions and the spectra of these molecules.
Conversely if the hadrons contain heavy quarks,
then heavy{-}quark symmetry in its different manifestations~\cite{Isgur:1989vq,Isgur:1989ed,Savage:1990di,Neubert:1993mb,Manohar:2000dt}
will influence the way these molecules organize in multiplets.
The light and heavy symmetries of hadronic molecules can indeed be used
to understand their known spectrum and to predict
the existence of new states~\cite{AlFiky:2005jd,Guo:2009id,Voloshin:2011qa,Mehen:2011yh,Valderrama:2012jv,Nieves:2012tt,HidalgoDuque:2012pq,Guo:2013sya,Guo:2013xga,Lu:2017dvm}.
The present manuscript deals with how these symmetries apply
for the particular case of a molecular $P_c(4450)$.

We will begin by considering the $P_c(4450)$ (the $P_c^*$ from now on)
from the point of view of heavy{-}quark spin symmetry (HQSS).
HQSS states that the dynamics of a heavy hadron is independent
of the spin of the heavy quark inside it.
In the molecular picture the $P_c^*$ is commonly pictured
as a $J^P = \frac{3}{2}^{-}$ $\bar{D}^* \Sigma_c$ molecule~\cite{Karliner:2015ina,Roca:2015dva,Xiao:2015fia,Burns:2015dwa,Geng:2017hxc},
or less commonly as a
$J^P = \frac{5}{2}^{-}$ $\bar{D}^* \Sigma_c^*$ molecule~\cite{Chen:2015loa}.
Here for concreteness we will work under the first of these assumptions,
namely that the $P_c^*$ is a $J^P = \frac{3}{2}^{-}$ $\bar{D}^* \Sigma_c$
bound state.
In either case we have to define heavy hadron superfields that
group the heavy hadron fields into units that are well-behaved
with respect to heavy{-}quark rotations.
The non-relativistic superfield for the heavy pseudoscalar and vector mesons $D$ and $D^*$ is~\cite{Falk:1992cx}
\begin{eqnarray}
  {H}_c = \frac{1}{\sqrt{2}}\,
  \left[ D + \vec{D}^* \cdot \vec{\sigma} \right] \, ,
\end{eqnarray}
where $H_c$ is a 2{$\times$}2 matrix and $\vec{\sigma}$ refers to the Pauli matrices.
For the heavy{-}baryon field we define the superfield as~\cite{Cho:1992cf}
\begin{eqnarray}
 \vec{S}_c = \frac{1}{\sqrt{3}}\,\vec{\sigma}\,\Sigma_c + \vec{\Sigma}_c^* \, ,
\end{eqnarray}
which is a 2{$\times$}3 matrix, basically the tensor product between
the spin-$1/2$ heavy and spin-$1$ light degrees of freedom.
In this representation the spin-3/2 heavy{-}baryon field is subjected
to the condition $\vec{\sigma} \cdot \vec{\Sigma}_c^* = 0$, which
ensures that the $\vec{\Sigma}_c^*$ is a spin-3/2 field.
From the heavy{-}meson and baryon superfield the most general contact-range
{L}agrangian that we can construct without derivatives is
\begin{eqnarray}
  \mathcal{L} &=& 
  C_a \,
  \vec{S}_c^{\dagger} \cdot \vec{S}_c\,
      {\rm Tr}\left[ {\bar H}_c^{\dagger} {\bar H}_c\right]
  \nonumber \\
  &+& C_b\,\sum_{i = 1}^3\,
  \vec{S}_c^{\dagger} \cdot (J_i \,\vec{S}_c)\,
      {\rm Tr}\left[ {\bar H}_c^{\dagger} \sigma_i {\bar H}_c\right] \, ,
\end{eqnarray}
where $J_i$ with $i = 1,2,3$ refers to the spin-1 angular momentum matrices,
which we recall here,
\begin{equation}
J_1 = \frac{1}{\sqrt{2}}\begin{pmatrix}
0 & 1 & 0 \\ 1 & 0 & 1 \\ 0 & 1 & 0
\end{pmatrix},\,\,\, J_2 = \frac{1}{\sqrt{2}}\begin{pmatrix}
0 & -i & 0 \\ i & 0 & -i \\ 0 & i & 0
\end{pmatrix},\,\,\, J_3 = \begin{pmatrix}
1 & 0 & 0 \\ 0 & 0 & 0 \\ 0 & 0 & -1
\end{pmatrix}, 
\end{equation}
and $C_a$ and $C_b$ are coupling constants.
This {L}agrangian leads to the contact-range potential
of Table \ref{tab:penta}, which in turn can be considered as
the leading{-}order potential of an effective field theory
for the $\bar{H}_c S_c$ family of molecules
(in line with the analogous effective field theories
for $H_c \bar{H}_c$~\cite{Valderrama:2012jv}
and $S_c \bar{S}_c$ molecules~\cite{Lu:2017dvm}).

\begin{table}[!ttt]
\begin{tabular}{|cccc|}
\hline \hline
  Molecule  & $J^{P}$ & $V$ & B (MeV) \\
  \hline
  $\bar{D} \Sigma_c$ & $\frac{1}{2}^-$ & $C_a$ & ? \\ \hline
  $\bar{D} \Sigma_c^*$ & $\frac{3}{2}^-$ & $C_a$ & ? \\ \hline
  $\bar{D}^* \Sigma_c$ & $\frac{1}{2}^-$ & $C_a - \frac{4}{3}\,C_b$ & ? \\
  $\bar{D}^* \Sigma_c$ & $\frac{3}{2}^-$ & $C_a + \frac{2}{3}\,C_b$ & $12 \pm 3$ \\
  \hline
  $\bar{D}^* \Sigma_c^*$ & $\frac{1}{2}^-$ & $C_a - \frac{5}{3}\,C_b$ & ? \\
  $\bar{D}^* \Sigma_c^*$ & $\frac{3}{2}^-$ & $C_a - \frac{2}{3}\,C_b$ & ? \\
  $\bar{D}^* \Sigma_c^*$ & $\frac{5}{2}^-$ & $C_a + C_b$ & $12$ \\
  \hline \hline 
\end{tabular}
\caption{The lowest{-}order contact-range potential
  for the $\bar{H}_c S_c$ system, which contains
  two unknown couplings $C_a$ and $C_b$.
  We show the potential for each particle and spin channel
  (the ``Molecule'' and ``$J^P$'' columns).
  The potential is suspected to bind for the $J^P = \frac{3}{2}^{-}$
  ${\bar D}^* \Sigma_c$ channel, forming the $P_c(4450)$ pentaquark,
  which strongly suggest{s} that the $J^P = \frac{5}{2}^{-}$
  ${\bar D}^* \Sigma_c^*$ channel binds too.
}
\label{tab:penta}
\end{table}

If the $P_c(4450)$ is a $\frac{3}{2}^-$ $\bar{D}^* \Sigma_c$ molecule,
its potential is
\begin{eqnarray}
  V (\bar{D}^* \Sigma_c, J^P = \tfrac{3}{2}^-) = C_a + \tfrac{2}{3}\,C_b \, .
\end{eqnarray}
Curiously, the potential for a prospective
$\frac{5}{2}^-$ $\bar{D}^* \Sigma_c^*$ molecule is similar:
\begin{eqnarray}
  V (\bar{D}^* \Sigma_c^*, J^P = \tfrac{5}{2}^-) = C_a + C_b \, .
\end{eqnarray}
which strongly suggest{s} that this molecule should also bind (probably
with a binding energy similar to that of the $\frac{3}{2}^{-}$ state).
This conclusion is subject to a series of uncertainties, from which
the most obvious one is that the potential is not exactly the same.
We do not know how much of the binding is due to the individual
couplings $C_a$ and $C_b$.
It could indeed happen that one of these states binds,
but the other does not, which could be the case if $|C_b|$
is disproportionately bigger than $|C_a|$.
The existence of the $\frac{5}{2}^{-}$ $\bar{D}^* \Sigma_c^*$ partner
will be very likely if the two couplings are of similar size,
i.e. $|C_a| \sim |C_b|$, or alternatively if $|C_a| > |C_b|$.
In this regard we note that the phenomenological model of
Ref.~\cite{Wu:2010jy} (which predicts $\bar{D} \Sigma_c$ and
$\bar{D}^* \Sigma_c$ bound states at $4261$ and $4412\,{\rm MeV}$
respectively, in the latter case independently of
the total spin of the $\bar{D}^* \Sigma_c$ system)
indeed suggest that $|C_a| > |C_b|$.
Besides the issue with $C_a$ and $C_b$, we have that HQSS is not exact
but{ }expected to have a level of uncertainty
of the order of $\Lambda_{\rm QCD} / m_Q$, with
$\Lambda_{\rm QCD} \sim 200-300\,{\rm MeV}$ and $m_Q$ the mass of the heavy quark.
For the charm sector this uncertainty is of the order of $15\%$, which
is how much the potential in the $\frac{5}{2}^{-}$ $\bar{D}^* \Sigma_c^*$
molecule is expected to differ from its HQSS expectation.
The existence of subleading{-}order effects, in particular one{-}pion exchange,
induces an additional source of uncertainty.
This is usually dealt with by including a floating {cutoff}
in the calculations and varying it within a reasonable window,
as we will show later for the triply heavy molecules.
Other effect from one{-}pion exchange is the curious coupled{-}channel dynamics
between the $\frac{3}{2}^-$ $\bar{D}^* \Sigma_c$ and
$\bar{D} \Lambda_{c}(2595)$ channels ($\bar{D} \Lambda_{c1}$ from now on).
This involves the exchange of a pion near the mass shell, resulting
in a long-range $1/r^2$ type of interaction that renders
binding easier~\cite{Geng:2017hxc}.
For the $\frac{3}{2}^-$ $\bar{D}^* \Sigma_c$-$\bar{D} \Lambda_{c1}$ system
this effect is modest, but still noticeable: the short-range attraction
(i.e. the $C_a + \tfrac{2}{3}\,C_b$ coupling combination)
required to bind the coupled
$\frac{3}{2}^-$ $\bar{D}^* \Sigma_c$-$\bar{D} \Lambda_{c1}$ system
is about $70-90\%$  of that required to bind the uncoupled
$\frac{3}{2}^-$ $\bar{D}^* \Sigma_c$ system, depending
on whether we include pions or not~\cite{Geng:2017hxc}.

For the doubly heavy baryons we define the superfield~\cite{Hu:2005gf}
\begin{eqnarray}
  \vec{T}_{cc} = \frac{1}{\sqrt{3}}\,\vec{\sigma}\,\Xi_{cc} + \vec{\Xi}_{cc}^* \, ,
\end{eqnarray}
which is formally analogous to the $\vec{S}_c$ superfield.
But the interpretation of the $\vec{T}_{cc}$ superfield is different
from the $\vec{S}_c$ superfield: for $\vec{T}_{cc}$ the light spin
is $1/2$ while the heavy spin is $1$.
The application of heavy{-}antiquark-diquark symmetry (HADS)~\cite{Savage:1990di} 
can actually be encapsulated in the following two substitutions
\begin{eqnarray}
  {\rm Tr}\left[ {\bar H}_c^{\dagger} {\bar H}_c\right] &\to&
  \vec{T}^{\dagger}_{cc} \cdot \vec{T}_{cc} \, , \\
      {\rm Tr}\left[ {\bar H}_c^{\dagger} \sigma_i {\bar H}_c\right] &\to&
    \vec{T}^{\dagger}_{cc} \cdot (\sigma_i \vec{T}_{cc}) \, ,
\end{eqnarray}
which are derived from the formalism of Ref.~\cite{Hu:2005gf}.
From these substitutions we arrive at the Lagrangian that describes
the contact-range interaction between a heavy baryon
and a doubly heavy baryon:
\begin{eqnarray}
  \mathcal{L} &=& 
  C_a \,
  \vec{S}_c^{\dagger} \cdot \vec{S}_c\,
  \vec{T}_{cc}^{\dagger} \cdot \vec{T}_{cc}\,
  \nonumber \\
  &+& C_b\,\sum_{i=1}^3
  \vec{S}_c^{\dagger} \cdot (J_i\, \vec{S}_c)\,
  \vec{T}_{cc}^{\dagger} \cdot (\sigma_i \vec{T}_{cc})\, ,
\end{eqnarray}
where the corresponding contact-range potential
can be consulted in Table \ref{tab:hexa}.
The following configurations are worth considering:
\begin{eqnarray}
  V (\Xi_{cc} \Sigma_c, J^P = 0^+) &=& C_a + \tfrac{2}{3}\,C_b \, , \\
  V (\Xi_{cc} \Sigma_c^*, J^P = 1^+) &=& C_a + \tfrac{5}{9}\,C_b \, , \\
  V (\Xi_{cc}^* \Sigma_c, J^P = 2^+) &=& C_a + \tfrac{2}{3}\,C_b \, , \\
  V (\Xi_{cc}^* \Sigma_c^*, J^P = 3^+) &=& C_a + C_b \, ,
\end{eqnarray}
because they imply a potential that is either identical
to that of a molecular {$P_c^*$} or very similar.
From this it is sensible to expect that in a first approximation
these four molecules will bind.
More concrete predictions are possible from solving a non-relativistic
bound state equation with the contact-range potentials.
If we work in momentum space, we can solve the integral equation
\begin{eqnarray}
  \phi(k) + \int \frac{d^3 p}{(2\pi)^3}\,\langle k | V | p \rangle
  \,\frac{\phi(p)}{B + \frac{p^2}{2 \mu}} = 0 \, ,
\end{eqnarray}
where $\phi$ is the vertex function,
$B$ the binding energy, and $\mu$ the reduced mass.
For solving this equation we have to regularize the contact-range potential
\begin{eqnarray}
  \langle p | V_{\Lambda} | p' \rangle =  C_{P^*_c}(\Lambda)\,
  f(\tfrac{p}{\Lambda})\,f(\tfrac{p'}{\Lambda}) \, ,
\end{eqnarray}
with $\Lambda$ a {cutoff}, $f(x)$ a regulator function and
$C_{P^*_c} = C_a + \frac{2}{3}\,C_b$ the coupling of
the contact-range potential for the $P_c^*$ and
the $0^+$ $\Xi_{cc} \Sigma_c$ and $2^+$ $\Xi_{cc}^* \Sigma_c$ molecules,
see Tables \ref{tab:penta} and \ref{tab:hexa}.
A typical choice of the {cutoff} is $\Lambda = 0.5-1.0\,{\rm GeV}$,
while for the regulator we will choose $f(x) = e^{-x^2}$.
For the masses we use $m(D^*) = 2009\,{\rm MeV}$,
$m(\Sigma_c) = 2454\,{\rm MeV}$, $m(\Sigma_c^*) = 2518\,{\rm MeV}$
(i.e. the isospin average of their PDG values~\cite{Tanabashi:2018oca}),
$m(\Xi_{cc}) = 3621\,{\rm MeV}$ and $m(\Xi_{cc}^*) = 3727\,{\rm MeV}$,
where the $\Xi_{cc}^*$ mass has been deduced from the HADS relation
$m(\Xi_{cc}^*) - m(\Xi_{cc}) = \frac{3}{4}\left(m(D^*) - m(D)\right)$~\cite{Savage:1990di}.

We can use the existence of the $P_c^*$ as a renormalization condition,
that is, for a given {cutoff} $\Lambda$ and regulator function
we fix the coupling $C_{P^*_c}(\Lambda)$ from the condition of
reproducing the $P_c^*$ pole.
With the coupling determined in this way, we can make predictions
for the $0^+$ and $2^+$ triply heavy molecules:
\begin{eqnarray}
  B(0^+) \simeq  B(2^+) \simeq 19-29\,{\rm MeV}\, , 
\end{eqnarray}
which are more bound than the original $P_c^*$ state simply
because the reduced mass is bigger for the heavy{-}baryon {/}
doubly{-}heavy{-}baryon system.
The above range represents the cutoff variation,
which is expected to {give} the uncertainty from not taking into account
subleading-order interactions such as pion exchanges
though we will comment on this later.
For the $1^+$ and $3^+$ molecules the binding energy should be similar,
but it is difficult to be more concrete as the potentials are not
exactly the same.
We mention in passing that phenomenological predictions of
$0^+$ $\Xi_{cc} \Sigma_c$ and $1^+$ $\Xi_{cc} \Sigma_c^*$ molecules also
exist in the one boson exchange model~\cite{Chen:2018pzd},
though they are more vague owing to the non-uniqueness of
physically acceptable form factor and cutoff choices.

\begin{table}[!h]
\begin{tabular}{|cccc|}
\hline \hline
  Molecule  & $J^{P}$ & $V$ & B (MeV) \\
  \hline
  $\Xi_{cc} \Sigma_c$ & $0^+$ & $C_a + \frac{2}{3}\,C_b$ &
  $19^{+15}_{-13}$($29^{+32}_{-23}$) \\
  $\Xi_{cc} \Sigma_c$ & $1^+$ & $C_a - \frac{2}{9}\,C_b$ & ? \\
  \hline
  $\Xi_{cc} \Sigma_c^*$ & $1^+$ & $C_a + \frac{5}{9}\,C_b$ & $20-30$ \\
  $\Xi_{cc} \Sigma_c^*$ & $2^+$ & $C_a - \frac{1}{3}\,C_b$ & ? \\
  \hline
  $\Xi_{cc}^* \Sigma_c$ & $1^+$ & $C_a - \frac{10}{9}\,C_b$ & ? \\
  $\Xi_{cc}^* \Sigma_c$ & $2^+$ & $C_a + \frac{2}{3}\,C_b$
  & $19^{+15}_{-12}$($30^{+33}_{-24}$) \\
  \hline
  $\Xi_{cc}^* \Sigma_c^*$ & $0^+$ & $C_a - \frac{5}{3}\,C_b$ & ? \\
  $\Xi_{cc}^* \Sigma_c^*$ & $1^+$ & $C_a - \frac{11}{9}\,C_b$ & ? \\
  $\Xi_{cc}^* \Sigma_c^*$ & $2^+$ & $C_a - \frac{1}{3}\,C_b$ & ? \\
  $\Xi_{cc}^* \Sigma_c^*$ & $3^+$ & $C_a + C_b$ & $20-30$ \\
  \hline \hline 
\end{tabular}
\caption{The lowest order contact-range potential for the $T_{cc} S_c$ system,
  which we derive from the $\bar{H}_c S_c$ potential and HADS.
  The potential depends on two unknown couplings $C_a$ and $C_b$, with
  different linear combinations depending on the particle
  and spin channel (the ``Molecule'' and ``$J^P$'' columns).
  The combination $C_a + \frac{2}{3} C_b$ can be determined from the hypothesis
  that the $P_c^*$ is a $\bar{H}_c S_c$ molecule. From this we can compute
  the binding energy of two $T_{cc} S_c$ molecules and estimate the binding
  energy of another two.
  The binding energies are expressed in ${\rm MeV}$. For the $0^+$ and $2^+$
  molecules we show the results for the {cutoff}
  $\Lambda = 0.5$ ($1.0$) ${\rm GeV}$,
  where the errors come from the uncertainty of HADS and the $P_c^*$ mass.
  For the $1^+$ and $3^+$ molecules we simply show the {cutoff} variation:
  the uncertainty in the binding of these states is difficult
  to estimate because the potential is in fact not identical
  to that of the $P_c^*$.
}
\label{tab:hexa}
\end{table}

Of course there are several sources of uncertainty
that have to be taken into account.
The most conspicuous one is the binding energy
of the $P_c^*$, i.e. $B = 12 \pm 3\,{\rm MeV}$.
A second source of uncertainty is HADS itself,
which is expected to be only accurate at the $\Lambda_{\text{QCD}} / (m_Q v)$
level~\cite{Savage:1990di} with $v$ the velocity of
the heavy diquark pair~\footnote{
  Ref.~\cite{Hu:2005gf} estimates $m_Q v \sim 0.8\,{\rm GeV}$
  for the case of the charm quark ($Q = c$),
  while Ref.~\cite{Cohen:2006jg} argues that the $c$ quark
  is too light for HADS to be applicable.
  From lattice QCD~\cite{Padmanath:2015jea}  it seems apparent
  that the $J=\frac{1}{2}$ $\Xi_{cc}$ and $J=\frac{3}{2}$ $\Xi_{cc}^*$
  mass splitting is close to the HADS prediction.
  The eventual discovery of the $\Xi_{cc}^*$ doubly charmed baryon
  and its properties will probably settle the question of how
  accurate is HADS.}.
This is easily included by assuming that the relative error of
the $0^+$ and $2^+$ potential is of the same size as the HADS uncertainty,
which we estimate to be about $30\%$.
Combining these two effects, we arrive at
\begin{eqnarray}
  B(0^+) \simeq  B(2^+) \simeq 19^{+15}_{-13} \, (29^{+33}_{-24})\,{\rm MeV} \, ,
  \label{eq:B0-p}
\end{eqnarray}
for $\Lambda = 0.5\,(1.0)\,{\rm GeV}$. For a more comprehensive list
we refer to Table \ref{tab:hexa}, where it should be noticed that
for the $1^+$ and $3^+$ states this detailed error analysis
is not possible because the potential does not exactly
match that of the $P_c^*$.
The errors are dominated by the HADS uncertainty, with the binding uncertainty
playing a secondary role. 
There are additional sources of uncertainly which are not so easily {modeled}.
One of these sources is the contribution of
the $\bar{D} \Lambda_{c1}$ channel to the binding of the $P_c^*$.
As already explained, this contribution basically reduces the strength of
the contact-range interaction required to bind the $P_c^*$.
However the error induced by this effect is expected to be noticeably
smaller than the HADS uncertainty.
One{-}pion exchange will also be another important factor, and so will be coupled{-}channel effects ($\bar{D}^* \Sigma_c$-$\bar{D}^* \Sigma_c^*$).
Yet these uncertainty sources fall into the category of
subleading{-}order contributions and are in principle
expected to be covered by the {cutoff} variation.
There is a caveat though: when connecting the double and triply heavy molecules,
calculations do not converge in the $\Lambda \to \infty$ limit
{(the binding energy eventually develops a quadratic divergence)}.
This is analogous to what happens when connecting different
heavy{-}flavour sectors~\cite{Baru:2018qkb}.
As a consequence the calculations cannot be interpreted
as the results of a genuine EFT, but instead
have a distinct phenomenological taste.
In practical terms this means that there are systematic errors
that have not been accounted for properly.
Luckily the size of these unaccounted errors seems to be moderate:
despite the impossibility of removing the {cutoff},
the predictions are relatively independent of
the choice of regulator (provided we use a {cutoff} of the order of
the typical hadronic scale).
This can be illustrated with the use of a delta-shell regulator
in coordinate space
\begin{eqnarray}
  V(r; R_c) = C_{P^*_c}(R_c)\,\frac{\delta( r- R_c)}{4\pi R_c^2} \, ,
\end{eqnarray}
with $R_c = 0.5-1.0\,{\rm fm}$ (i.e. the typical size of hadrons).
With this regulator the predictions are
\begin{eqnarray}
  B(0^+) \simeq  B(2^+) \simeq 20^{+18}_{-14} \, (31^{+37}_{-25})\,{\rm MeV} \, ,
  \label{eq:B0-r}
\end{eqnarray}
for $R_c = 1.0\,(0.5)\,{\rm fm}$, which is fairly consistent with the previous
results with the {Gaussian} 
regulator in momentum space, see Eq.~(\ref{eq:B0-p}).

Be it as it may, the biggest source of systematic uncertainty is the fact
that we do not know the nature of the $P_c^*$ for sure.
Here we have simply followed the hypothesis
that it is a $\frac{3}{2}^-$ $\bar{D}^* \Sigma_c$ molecule~\cite{Karliner:2015ina,Roca:2015dva,Xiao:2015fia,Burns:2015dwa,Geng:2017hxc}
with a binding energy of about $12\,{\rm MeV}$ and a radius of
$1/\!\sqrt{2 \mu B} \sim 1.2\,{\rm fm}$.
Other authors consider the $P_c^*$ to be a
$\frac{5}{2}^{-}$ $\bar{D}^* \Sigma_c^*$
molecule~\cite{Chen:2015loa} with a binding energy of about $77\,{\rm MeV}$,
implying a radius of $0.5\,{\rm fm}$.
In this second scenario, the $P_c^*$ is considerably more compact to the {extent}
that it is somewhere in the limit between a molecule and a multiquark state.
We will not consider this scenario in detail here, but merely mention that the
qualitative predictions will not change: we still expect a $\frac{3}{2}^{-}$
pentaquark partner and the $0^+$, $1^+$, $2^+$ and $3^+$ $T_{cc} S_c$ molecules.
The quantitative predictions are fairly different though.
The $\frac{3}{2}^{-}$ pentaquark will be bound by about $70-80\,{\rm MeV}$
and will be located in the $4380\,{\rm MeV}$ region,
where the second LHCb pentaquark is.
Their identification is problematic, in part owing to the large
decay width of the $P_c(4380)$ (mostly into $J/\psi \, p$),
$\Gamma = 205 \pm 90\,{\rm MeV}$, which is definitely
large for a composed state, but not that excessive
if we take into account that the size of the $P_c(4380)$
in this scenario is similar to the size of the $J/\psi$,
namely $0.47\,{\rm fm}$~\cite{Eichten:1978tg}.
Actually the strongest argument against the $P_c(4380)$ being the HQSS
partner of the $P_c(4450)$ comes from the statistical analysis of
the LHCb data, from which it is unlikely that the two pentaquark-like
resonances have the same parity~\cite{Jurik:2206806}.
For the $3^+$ $\Xi_{cc}^* \Sigma_c^*$ molecule,
the binding energy will be $B(3^+) \simeq 90-110\,{\rm MeV}$,
with similar binding energies expected
for its $0^+$, $1^+$ and $2^+$ partners.
Finally there are a series of works that do not consider the $P_c^*$
to be a molecule: the predictions of this manuscript are not likely
to hold if these scenarios are confirmed. But this depends on whether
it is the dynamics of the $P_c^*$ or heavy quark symmetry itself
which plays {the greatest} role in the eventual existence of
these partners.

To summarize, we have explored what consequences can be derived from
HQSS and HADS and the hypothesis that the $P_c^*$ is a molecular state.
From HQSS it is plausible to expect that the $P_c^*$ has a $\frac{5}{2}^-$
partner, with a similar binding energy to that of the $P_c^*$,
which we may call the $P_c(4515)$.
From HADS we expect the existence of up to four triply heavy baryon-baryon
molecules, with quantum numbers $0^+$, $1^+$, $2^+$ and $3^+$ and
binding energies in the $20-30\,{\rm MeV}$ range.
These predictions are subjected to a series of uncertainties,
which include the approximate nature of HQSS and HADS,
the existence of a long-range $1/r^2$ potential
in the $P_c^*$ molecular candidate,
subleading contributions to the potential such as pion exchanges,
and the fact that in a few instances the form of the potential is
not identical to that of the $P_c^*$.
The most important systematic uncertainty is the nature of the $P_c^*$ itself,
which will require further experiments.
While the eventual experimental observation of the theorized HQSS partner of
the $P_c^*$ seems possible, the detection of its triply charmed partners
is more tricky as it requires triple charm production.
This suggests that the lattice might be a more expedient way for determining
the existence of $\Xi_{cc} \Sigma_c$, $\Xi_{cc} \Sigma_c^*$,
$\Xi_{cc}^* \Sigma_c$ and $\Xi_{cc}^* \Sigma_c^*$ molecules.

\section*{Acknowledgments}
We thank Li-Sheng Geng and Feng-Kun Guo for comments,
suggesting references and a careful reading of the manuscript.
We also thank Eulogio Oset for clarifications and discussions.
This work is partly supported by the National Natural Science Foundation of
China under Grants No.11522539, 11735003, the fundamental Research Funds
for the Central Universities and the Thousand
Talents Plan for Young Professionals.

\appendix


%

\end{document}